%% file: dom_firewall.tex
\documentclass[letterpaper,twocolumn,10pt]{article}
\usepackage{usenix2019_v3}

\usepackage{tikz}
\usepackage{amsmath}

\usepackage{filecontents}

\usepackage{booktabs} 
\usepackage{listings}
\usepackage{url}
\usepackage{caption}
\usepackage{subcaption}
\usepackage[ruled,linesnumbered]{algorithm2e}
\usepackage[printonlyused,nohyperlinks]{acronym}

\acrodef{xhr}[XHR]{{\tt XMLHttpRequest}}
\acrodef{WAF}[WAF]{Web Application Firewall}
\acrodef{XSS}[XSS]{Cross-Site Scripting}
\acrodef{CSRF}[CSRF]{Cross-Site Request Forgery}
\acrodef{CSP}[CSP]{Content Security Policy}
\acrodef{CVE}[CVE]{Common Vulnerability and Exposures}
\acrodef{DDoS}[DDoS]{Distributed Denial-of-Service}
\acrodef{SQL}[SQL]{Structured Query Language}
\acrodef{CMS}[CMS]{Content Management System}
\acrodef{PoC}[PoC]{proof of concept}

\definecolor{lightgray}{rgb}{.9,.9,.9}
\definecolor{darkgray}{rgb}{.4,.4,.4}
\definecolor{purple}{rgb}{0.65, 0.12, 0.82}
\lstdefinelanguage{JavaScript}{
  keywords={typeof, new, true, false, catch, function, return, null, catch, switch, var, if, in, while, do, else, case, break},
  keywordstyle=\color{black}\bfseries,
  ndkeywords={export, boolean, throw, implements, import, this}, 
  ndkeywordstyle=\color{darkgray}\bfseries,
  identifierstyle=\color{black},
  sensitive=false,
  comment=[l]{//},
  morecomment=[s]{/*}{*/},
  commentstyle=\color{black}\ttfamily,
  stringstyle=\color{black}\ttfamily,
  morestring=[b]',
  morestring=[b]"
}
\newcommand{\sys}[0]{XSnare\xspace}
\newcommand{\xss}[0]{\ac{XSS}\xspace}

\newcommand{\js}[1]{\lstinline[mathescape,basicstyle=\normalsize]^#1^}
\newcommand{\code}[1]{\lstinline[mathescape,basicstyle=\normalsize]^#1^}

\input{annotations}

\begin{document}
	
\date{}

\title{\Large Precise XSS detection and mitigation with Client-side Templates}

\author{
{\rm Jos\'e Carlos Pazos}\\
University of British Columbia
\and
{\rm Jean-S\'ebastien L\'egar\'e}\\
University of British Columbia
\and
{\rm Ivan Beschastnikh}\\
University of British Columbia
\and
{\rm William Aiello\thanks{Dr. Aiello has provided crucial expert advice and insight in the early stages of the project. We miss him dearly.}}\\
University of British Columbia
}
\maketitle

\input{abstract}

\acresetall	

\acused{CVE}
\acused{SQL}
\acused{DDoS}

\input{introduction}
\input{client_approach}

\input{implementation}
\input{signature_development}

\input{methodology}

\input{performance}
\input{limitations}
\input{related_work}

\input{conclusion}

\bibliographystyle{plain}
\bibliography{references}

\input{appendix}

\end{document}

%% file: annotations.tex
\usepackage{xspace}
\usepackage{tikz}
\usepackage{array}
\usepackage{xcolor}

\newcolumntype{C}[1]{>{\centering\let\newline\\\arraybackslash}m{#1}}

\newboolean{showcomments}

\setboolean{showcomments}{true}

\ifthenelse{\boolean{showcomments}}
{\newcommand{\nbc}[3]{
 {\colorbox{#3}{\bfseries\sffamily\scriptsize\textcolor{white}{#1}}}
 {\textcolor{#3}{\sf\small$\blacktriangleright$\textit{#2}$\blacktriangleleft$}}}
}{\newcommand{\nbc}[3]{}
 }

\definecolor{ibcolor}{rgb}{0.4,0.6,0.2}
\definecolor{jpcolor}{rgb}{0.2,0.6,0.6}
\definecolor{nkcolor}{rgb}{0.8,0.5,0.3}
\definecolor{sbcolor}{rgb}{0.0,0.0,1}

\definecolor{todocolor}{rgb}{0.9,0.1,0.1}

%% file: abstract.tex
\begin{abstract}

We present \sys, a fully client-side \ac{XSS} solution, implemented as
a Firefox extension. Our approach takes advantage of available
previous knowledge of a web application's HTML template content, as
well as the rich context available in the DOM to block XSS
attacks. \sys prevents \ac{XSS} exploits by using a database of
exploit descriptions, which are written with the help of previously
recorded CVEs. CVEs for XSS are widely available and are one of the
main ways to tackle zero-day exploits. \sys effectively singles out
potential injection points for exploits in the HTML and sanitizes
content to prevent malicious payloads from appearing in the DOM.

\sys can protect application users
before application developers release patches and before server
operators apply them.

We evaluated \sys on 81 recent CVEs related to XSS
attacks, and found that it defends against 94.2\% of these
exploits. To the best of our knowledge, \sys is the first protection
mechanism for XSS that is application-specific, and based on publicly
available CVE information.  We show that \sys{}'s specificity protects
users against exploits which evade other, more generic, anti-XSS
approaches.

Our performance evaluation shows that our extension's overhead on web
page loading time is less than 10\% for 72.6\% of the sites in the Moz
Top 500 \cite{top500} list.

\end{abstract}

%% file: introduction.tex
\section{Introduction} \label{introduction}

\ac{XSS} is still one of the most dominant web vulnerabilities. A 2017
report showed that 50\% of websites contained at least one \ac{XSS}
vulnerability \cite{Acunetix}. Countermeasures exist, but many of them
lack widespread deployment, and so web users are still mostly
unprotected.

Informally, the cause of \xss is a lack of input sanitization:
user-chosen data ``escapes'' into a page's template and makes its way
into the JavaScript engine, or modifies the DOM.
Consequently, many of the \xss defenses published so far
propose to fix the problem at the source, by properly separating the
template from the user data on the server, or by modifying
browsers~\cite{Jim:2007:DSI:1242572.1242654,Nadji:2009,Wurzinger:2009:SMX:1656360.1656379,Sundareswaran:2012:XHS:2352970.2352994}.
There are also similar solutions that can be implemented in the
front-end code of an application~\cite{10.1007/978-3-319-66399-9_7}.
In all cases, these technologies must be adopted by the application
software developers, otherwise users are left unprotected.

One barrier to adoption of existing \xss defenses is that developers may not have the
necessary expertise, or sufficient resources, to use the approach.
%
%
Luckily, users wishing to gain reassurance over the safety of the
sites they visit can install browser extensions to filter malicious
scripts and content. Unfortunately, these extensions achieve most of
their security by disabling functionality in the applications, such as JavaScript, which
impairs the usability of the
sites~\cite{Noscript,Snyder:2017:MWD:3133956.3133966}. For example, most sites rely
on JavaScript being enabled\footnote{As early as 2012, it is used by almost
  100\% of the Alexa top 500
  sites~\cite{Stock:2017:WTI:3241189.3241265}}. 

When an \xss vulnerability is disclosed, some software vendors respond
with patches. If the affected software is released in the form of
packages, frameworks, or libraries, and used by several web
applications, there is delay before users can benefit from the
patch. Most importantly, the patched software must be re-deployed by
site administrators.

Unfortunately, website administrators will not, and often cannot,
apply software updates immediately: one study found
that 61\% of WordPress websites were running a version with known
security vulnerabilities~\cite{Sucuri}. In another report, we learn
that 30.95\% of Alexa's top 1 Million sites run a vulnerable version
of WordPress~\cite{wpwhitesecurity}.

Users are in effect at the mercy of developers and administrators if
they need to browse safe, up-to-date, applications. Our solution, \textbf{\sys},
helps with this problem -- based on information from past disclosures,
\sys patches known page vulnerabilities directly in the browser.

\begin{figure}[h]
  \includegraphics[scale=0.37]{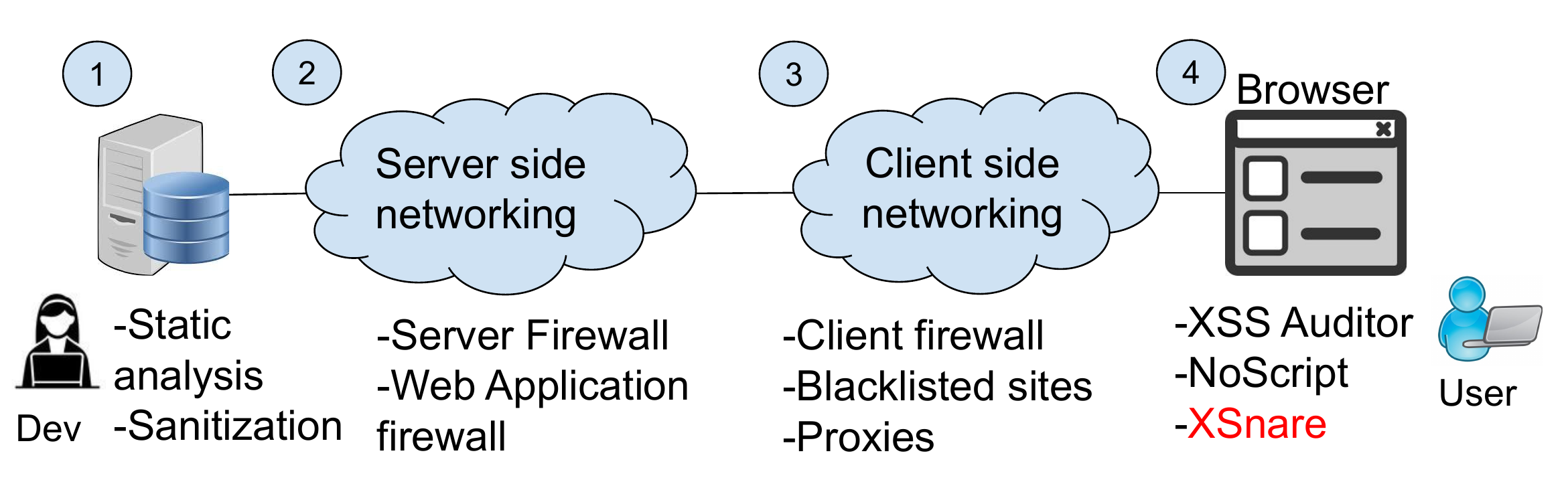}
  \vspace*{-5.0ex}
  \caption{Different web security solutions with \sys on the client-side.}
  \label{fig:web_architecture}
\end{figure}

Each layer of the web application stack  (Figure~\ref{fig:web_architecture}) opens different defence options against \xss:
\begin{enumerate}

\item The application logic is the first line of defence.
  Code safety can be enhanced with third-party vulnerability scanning solutions, and a thorough
  code-review process. Taint, and static code analysis tools can detect unsanitized inputs.

\item In the hosting environment, network firewalls, specifically \acp{WAF} can defend against attacks such as \ac{DDoS}, \ac{SQL} injections and \ac{XSS}.

\item In the client's environment (residential or commercial), users may install network firewalls, network content filters, and web proxies.

\item The last line of defence is the browser.
  Browser have built-in defences, such as
  Chrome's \ac{XSS} Auditor~\cite{xssauditor}. Users can also
  install third-party extensions to block malicious requests and
  responses, such as NoScript~\cite{Noscript}, and \sys.
\end{enumerate}

We make two observations about existing solutions: (a) server-side
solutions have to be applied independently on each server, and (b)
solutions on the client are typically written as generic filters which
attempt to catch everything, and consequently do not take full advantage
of the specificity of the application or the vulnerability.

For example, a \ac{WAF} can effectively protect the deployment placed
behind it, but users cannot realistically expect that every site they
visit be protected by one. At the opposite end, in the client's
environment, a user might configure a network proxy for all website
traffic, with generic rules achieving maximum coverage, but this
will often lead to an elevated rate of false positives (FPs).

Similarly, browser built-in defences are coarse-grained, and 
work on just a subset of exploits. Chrome's XSS Auditor, for example, only
attempts to defend against reflected \ac{XSS}. Google recently
announced its intention to deprecate XSS Auditor, for reasons
including \emph{``Bypasses abound''}, \emph{``It prevents some legit
  sites from working''}, and \emph{``Once detected, there's nothing
  good to do''}~\cite{deprecatexssauditor}. Stock et
al.~\cite{precise_dom_xss} propose enhancements to XSS Auditor and
cover a wider range of exploits than the auditor, but are limited to
DOM-based \ac{XSS}.  \textbf{By contrast, our work covers all types of \ac{XSS}.}

Implementing adequate server-side protections~\cite{Xu:2006:TPE:1267336.1267345,DBLP:conf/sec/Nguyen-TuongGGSE05,Pietraszek:2005:DAI:2146257.2146267,Bisht:2008:XPD:1428322.1428325} throughout a codebase could arguably qualify as a colossal task, considering the high turnaround times for resolving simpler bugs. A 2018 study found that the average time to patch a \ac{CVE}, all severities combined, is 38 days, increasing to as much as 54 days for low severity CVEs, and the oldest unpatched \ac{CVE} was 340 days old~\cite{Rapid7}.



Server-side defences also do not protect against client-only forms of
\xss, e.g., reflected \ac{XSS}, or persistent
client-side \ac{XSS}, which use a browser's local storage or cookies
as an attack vector. Steffens et
al.~\cite{DBLP:conf/ndss/SteffensRJS19} present a study of persistent
client-side \ac{XSS} across popular websites and find that as many as
21\% of the most frequented web sites are vulnerable to these attacks.
%
\textbf{To provide users with the means to protect themselves in the absence
of control over servers, we strongly believe that a novel client-side
solution is necessary.}

A number of existing solutions in this area also suffer from high
rates of false-positives and false-negatives. 
For example, NoScript~\cite{Noscript} works via domain white-listing, thus by
default, JavaScript scripts and other code will not execute. However,
not all scripts outside of the whitelist should be assumed to be
malicious. Browser-level filters like XSS Auditor work based on
general policies and can therefore incorrectly sanitize non-malicious
content.

\textbf{We posit that the DOM is the right place to mitigate XSS
  attacks as it provides a full picture of the web application.} While
most of the functionality we provide could be done by a network filter
in front of the browser, we take advantage of additional context
provided by the browser.
Particularly, when an exploit occurs as a result of user interactions,
like on response to a click, we
benefit from knowledge of the initiating tab to filter the
response. Previous client-side solutions have opted for detectors that were generic and site-agnostic~\cite{Kirda:2009:CCS:2639535.2639808,Jim:2007:DSI:1242572.1242654,Hallaraker:2005:DMJ:1078029.1078861}. Our work goes in the opposite direction, and tries rather to prevent precisely-defined exploits in specific applications.

If a patch for a server-side vulnerability can be ``translated'' into
an equivalent set of operations to apply on the fully formed HTML
document in the browser, then we can seize the opportunity to defend
{\em early} against exploits of that vulnerability.
Our extension, which has access to the user's browsing context, can
identify vulnerable pages based on a database of signatures for
previous disclosures. This way, \sys can protect users as soon as a patch
is implemented and added to its database. The client-side
patch will remain beneficial until \textit{all} server
operators running that software have had a chance to upgrade their
deployments.

 A similar philosophy is adopted by the client-side firewall-based network proxy Noxes~\cite{Kirda:2009:CCS:2639535.2639808}. 
Unlike \sys, however, Noxes only applies generic policies based on information available at the network layer. Namely, it does not protect against
attacks invisible to the network, e.g., deleting local files.
We believe additional contextual knowledge also offers more accurate vulnerability detection.




We evaluate \sys by
testing it on 81 recent \ac{XSS} CVEs. 
%
We also report \sys{}'s performance overhead on page load times
across a wide range of sites and show that it does not significantly
impact browsing experience.

To summarize, our contributions include:
\begin{itemize}

	\item \sys: a novel client-side framework that protects
          users against XSS vulnerabilities with a database of
          signatures for these vulnerabilities, written in a
          declarative language.

	\item A mechanism to correctly isolate a vulnerable injection
          point in a web page and to apply the intended server-side
          patch on the client-side.

	\item A collection of signatures to protect users against
          real XSS CVEs (\autoref{viability}), demonstrating the practicality
          of \sys; and the evaluation of its impact on browsing (\autoref{performance}). 

\end{itemize}

%% file: client_approach.tex
\section{\sys Design} \label{xsnare_design}

 \begin{figure}[h]
	\includegraphics[scale=0.55]{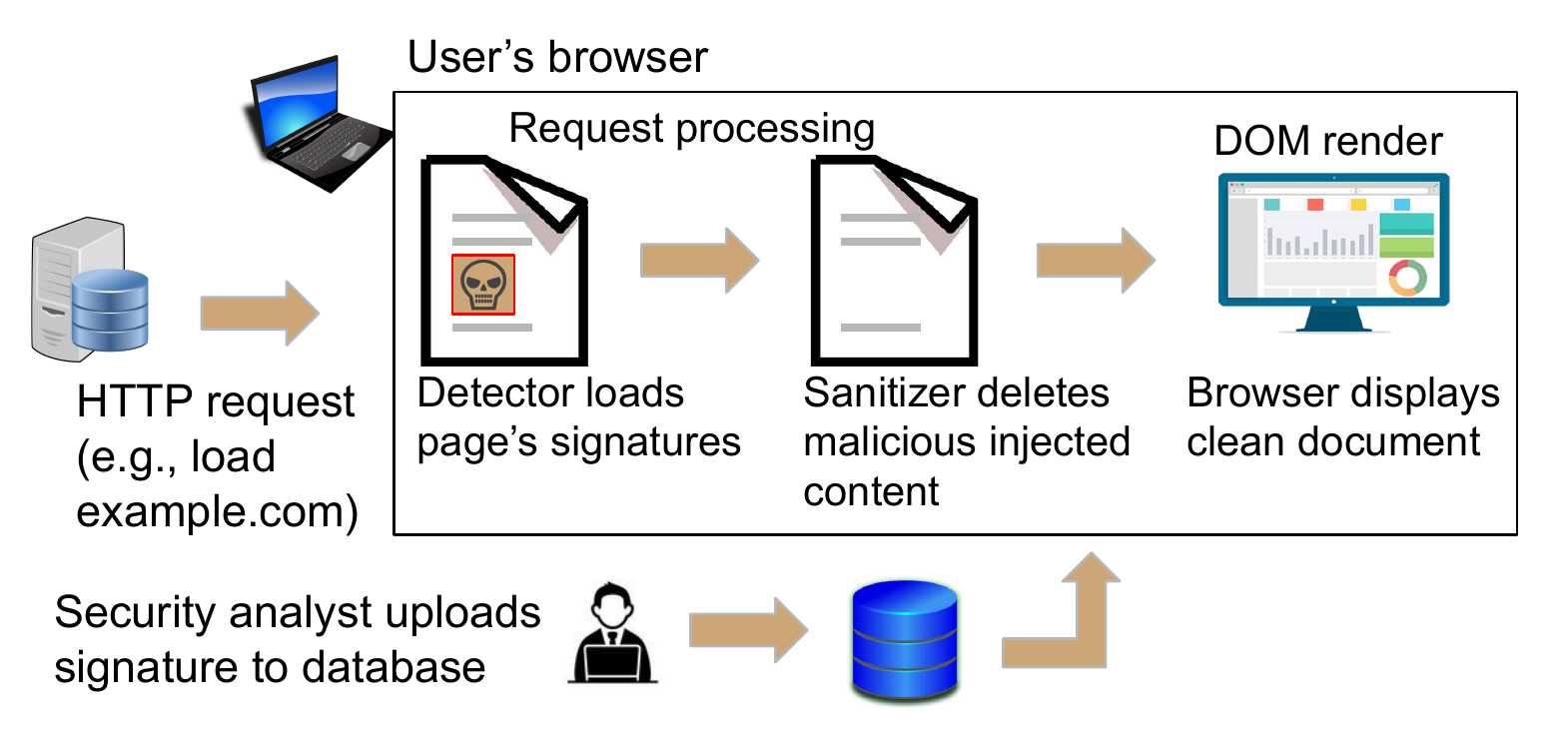}
	\caption{\sys's approach to protect against XSS.}
	\label{fig:xsnare}
\end{figure}

We now present the design of \sys and its components. 
%
%
We begin with a high-level view of its operation (see
\autoref{fig:xsnare}): A user requests a page, \url{example.com}, on a
browser with the \sys extension installed.
The response may or may not contain
malicious \xss payloads.
Before the browser renders the document, \sys analyzes
the potentially malicious document. The extension loads signatures
from its local database into its detector. The detector analyzes the
HTML string arriving from the network, and identifies the signatures
which apply to the document. These signatures specify one or more
``injection spots'' in the document, which correspond, roughly
speaking, to regions of the DOM where improperly sanitized content
could be injected.  The extension's sanitizer
eliminates any malicious content and outputs a clean HTML document to
the browser for rendering (\autoref{filter_algorithm}).


\subsection{An example application of \sys} \label{motivating_example}

To further explain our approach, we present a small example of how
DOM context can be used to defend against XSS, taken from CVE
2018-10309~\cite{examplecve}. This is reproducible in an off-the-shelf
WordPress installation running the Responsive Cookie Consent plugin,
v1.7. This is a stored \xss vulnerability, and as such is not caught
by some generic client-side \xss filters, including Chrome's XSS auditor.

Consider a website running PHP on the backend which stores user input
from one user, and displays it later to another user, inside an \textbf{input} element.

The PHP code defines the static HTML template (in black), as well as the dynamic input (in red):
\vspace{-0.2cm}
\begin{lstlisting}
<input id="rcc_settings[border-size]" 
name="rcc-settings[border-size]" 
type="text" value=<@\textcolor{red}{"<?php rcc\_value('border-size'); ?>"}@>/>
<label class="description"
for="rcc_settings[border-size]">
\end{lstlisting}
Normally, the \textbf{input} might have a value of "0":
\begin{lstlisting}
<input id="rcc_settings[border-size]" 
 name="rcc-settings[border-size]" 
 type="text" value=<@\textcolor{red}{"0"}@>>
<label class="description"
 for="rcc_settings[border-size]">
\end{lstlisting}
However, the php code is vulnerable to an injection attack, e.g.:
\begin{lstlisting}
border-size = ""><script>alert('XSS')</script>
\end{lstlisting}
The browser will render this, executing the injected script:
\begin{lstlisting}
<input id="rcc_settings[border-size]" 
name="rcc-settings[border-size]" 
type="text" value=<@\textcolor{red}{""><script>alert('XSS')</script>}@>
<label class="description"
for="rcc_settings[border-size]">
\end{lstlisting}

Note that the resulting HTML is well-formed, so a mere syntactic check
will not detect the malicious injection. Let us assume a security
analyst knows the original template, i.e., without injected
content. If the analyst were given a filled-in document, they could
(in most cases) separate the injected content from the server-side
template, and get rid of the malicious script entirely, using proper sanitization. 

The injected script is bounded by template elements with identifiable
attributes. Assuming (for now) that there is only one such vulnerable
injection point, we can search for the \textbf{input} element from the
top of the document, and the \textbf{label} from the bottom to ensnare
the injection points in the HTML.

This shares goals with the client/server hybrid approach of Nadji et
al.~\cite{Nadji:2009}. They automatically tag injected DOM elements on the
server-side using a taint-tracking, so that the client (a
modified browser) can reliably separate template vs
injected content. We do not require any server-side modifications, but rather opt for a client-side tagging solution based on exploit definitions.

The injected content, once identified, must be sanitized appropriately.
The appropriate action will depend on the application setting, but
assuming a patch has been written, it suffices to translate the
intention in the server code's path to the client-side.
This can be straightforward, once the fix is understood.

The developer incorrectly claimed the bug had
been fixed in version 1.8 of the plugin.
Other similar vulnerabilities had indeed been fixed, but not this one~\cite{rccpatch}. The built-in WordPress function \code{sanitize\_text\_field} needed to be applied.

\sys does not automatically determine the relevant actions to
implement from a patch. We assign this task to a security
analyst, who will act as the signature developer for a given
exploit. The system will however automate the signature matching and
sanitization.


\subsection{\sys Signatures} \label{signatures}



Our signature definitions make two assumptions: first,
\textbf{an injection must have a start point and end point}, that is,
an element can only be injected between a specific HTML node and its
immediate sibling in the DOM tree; second, in a well-formed DOM,
\textbf{the dynamic content will not be able to rearrange its location
  in the document without JavaScript execution} (e.g., removing and
adding elements), allowing us to isolate it from the template.

Pages commonly contain more than one vulnerable injection point.
We discuss the difficulty of supporting these pages in \autoref{multiple_injections}.

We believe CVEs are an ideal growing source of signature
definitions. Since previous client-side work does not focus on
application-specific protection, these tools often use less accurate
heuristics to detecting exploits. Furthermore, once new
vulnerabilities are found, these systems often lack the
maintainability obtained by leveraging active CVE development.

We are conscious that \sys signatures will not write themselves, and that
this task represents a new step in the workflow. Luckily, converting
the CVE information into a signature does not require active
participation from the application developers -- Security enthusiasts and
web developers have the skills to fulfill this role satisfactorily.

In general, we do not require the existence of a publicly disclosed CVE to be able to write a signature for an exploit, it is the process of its development that is useful to our approach (documenting an exploit and its cause). As described in \autoref{viability}, CVEs are a convenient way for us to test our system against real-world vulnerabilities. However, a knowledgeable analyst can write a signature without having publicly disclosed a CVE. In fact, for security measures, many CVEs are not publicly available until the application developer has patched its software. Our system can help defend against zero day attacks, as once a vulnerability is known, an analyst is able to write a signature for it as soon as they have knowledge of the issue. 

Long term, we imagine that volunteers (or entrepreneurs) would
cultivate and maintain the signature database. New signatures could be
contributed by a community of amateur or professional security analysts, in a manner not
so different from how antispam or antivirus software is managed.

The challenge of automatically deriving signatures from detailed CVEs is an interesting
one, albeit outside the scope of this paper.


 \subsection{Firewall Signature Language} \label{signature_language}

 Our signature language needs to be such that it has enough power of
 expression for the signature writer to be precise, both for
 determining the correct web application and to identify the affected
 areas in the HTML. For injection point isolation, a language based on
 regular expressions suffices to express precise
 sections of the HTML. The following is the signature that defends
 against the motivating example of \autoref{motivating_example}:

 \begin{lstlisting}[breaklines=true,caption={An \sys signature},label={lst:xsnare_signature}]
url: 'wp-admin/options-general.php?page=rcc-settings',
software: 'WordPress',
softwareDetails: 'responsive-cookie-consent',
version: '1.5',
type: 'string',
typeDet: 'single-unique',
sanitizer: 'regex',
config: '/^[0-9](\.[0-9]+)?$/',
endPoints: 
['<input id="rcc_settings[border-size]" name="rcc_settings[border-size]" type="text"
  value="',
'<label class="description" 
for="rcc_settings[border-size]">']
\end{lstlisting}

A description of the development process for this signature is given
in Section \ref{case_study}. In summary, a signature will have the
necessary information to determine whether a loaded page has a
vulnerability, and specify appropriate actions for eliminating
any malicious payloads.

Analysts configure their signatures with one
function chosen from the static set of sanitization
functions offered by \sys. These functions inoculate potentially malicious injections
based on the DOM context surrounding the injection. The goal of
signatures is to provide such sanitization, ideally without ``breaking''
the user experience of the page. The default function preset is DOMPurify's~\cite{10.1007/978-3-319-66399-9_7} default
configuration,
which takes care of common sanitization needs~\cite{safecontent}. However, DOMPurify's defaults can be unnecessarily restrictive, in which cases the other sanitization methods are more desirable.

We considered allowing arbitrary sanitization code in
signatures. While it would open complex sanitization possibilities, we
have decided against it, principally for security reasons. The minimal
set of functions we settled on also sufficed to express all of the
signatures defined for this paper. See ~\autoref{appendix:language_specification} for
more details.


 \subsection{Browser Extension} \label{firefox_extension}

 Our system's main component is a browser extension which rewrites
 potentially infected HTML into a clean document.
 The extension detects exploits in the HTML by using signature definitions and
 maintains a local database of signatures. We leave the design of an
 update mechanism to future work, but in its current form, the
 database is bundled with each new installation of the extension.

 The extension translates signature definitions into patches that
 rewrite incoming HTML on a per-URL basis, according to the top-down,
 bottom-up scan described in \autoref{motivating_example}.
 
 The extension's detector acts as an in-network filter. We initially considered other designs but quickly found out that applying the patch at the network level was necessary for sanitization correctness: even before any code runs, parsing the HTML into a DOM tree
might cause elements to be re-arranged into an unexpected order,
making our extension sanitize the wrong spot.  Consider the following
example, where an element inside a <tr> tag is rearranged after
parsing the string:
\vspace{-0.2cm}
\begin{lstlisting}
<table class="wp-list-table">
  <thead>
     <tr>
	     <th></th>
	     <@\textcolor{red}{<img src="1" onerror="alert(1)">}@>
	     <th>
   	     <form method="GET" action=""> ...
\end{lstlisting}

In this HTML, the signature developer might identify the exploit as
occurring inside the given table. However, if we wait until the string
has been parsed into a DOM tree to sanitize, the elements are
rearranged due to <tr> not allowing an <img> as its child:
\vspace{-0.2cm}
\begin{lstlisting}
<@\textcolor{red}{<img src="1" onerror="alert(1)">}@>
<table class="wp-list-table">
   <thead>
   <tr>
	   <th></th>
	   <th>
       <form method="GET" action=""> ...
\end{lstlisting}

Note that the injected <img> tag is now outside of the table, simply
by virtue of the DOM parsing. Now, the extension will search past the
injection, as it occurs before the table element, creating a false
negative (FN). Similarly, elements rearranged inside an injection
point can create false positives. This example would generate a class
of circumvention techniques for our detector, so we can't wait until
the website has been rendered to analyze the response. This guarantees
that a knowledgeable attacker can not take advantage of this
behavior.

\subsection{Handling multiple injections in one page} \label{multiple_injections}

In \autoref{lst:xsnare_signature}, the endPoints were listed as
two strings in the incoming network response. However, there are cases
where arbitrarily many injection points can be generated by the
application code, such as a for loop generating table rows. For these,
it is hard to correctly isolate each endPoint pair, as an attacker
could easily inject fake endPoints in between the original ones.

\begin{figure}[h]
	\begin{center}
	\includegraphics[scale=0.25]{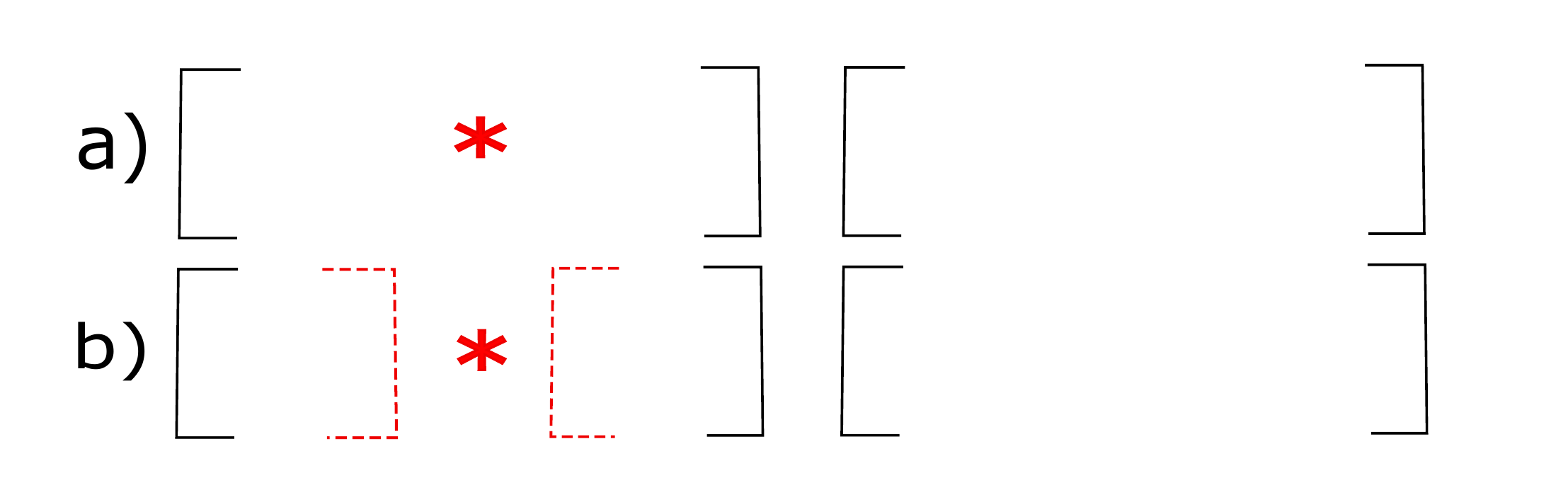}
	\caption{Example attacker injection when multiple injection points exist in the page. a) a basic injection pattern. b) an attempt to fool the detector.}
	\label{fig:attacker_injection}
	\end{center}
\end{figure}

In Figure~\ref{fig:attacker_injection}a, the brackets indicate a
template. The content in between is an injection point (the star),
where dynamic content is injected into the template. In the case of a
vulnerability, the injected content can expand to any arbitrary
string. The signature separates the injection from the rest by
matching for the start and end points (the \code{endPoints}),
represented by the brackets. This HTML originally has two pairs of
\code{endPoint} patterns.

In Figure~\ref{fig:attacker_injection}b, the attacker knows these are
being used as injection end points and decides to inject a fake ending
point and a fake starting point (the dotted brackets), with some
additional malicious content in between. If just looking for multiple
pairs of end points, the detector cannot tell the difference between
the solid and dotted patterns, and will not get rid of the content
injected in the star. Therefore, we have to use the first starting
point and the last ending point before a starting one (when searching
from the bottom-up) and sanitize everything in between. This might get
rid of a substantial amount of valid HTML, so we defer to the
signature developer's judgment of what behavior the detector should
follow. We expand upon this further in \autoref{case_study}.

\begin{figure}[h]
	\begin{center}
	\includegraphics[scale=0.25]{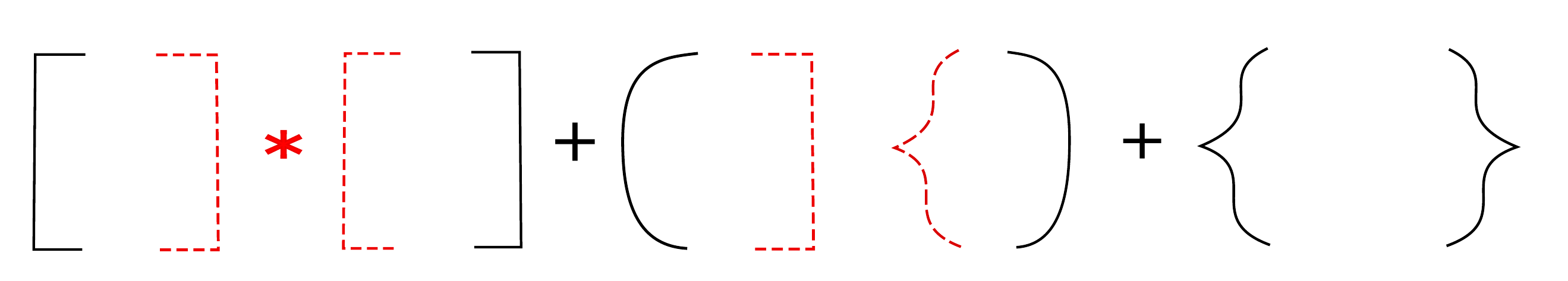}
	\caption{Example attacker injection when multiple distinct injection points exist in the page.}
	\label{fig:attacker_injection_unique}
	\end{center}
\end{figure}

~\autoref{fig:attacker_injection_unique} illustrates a case when
there are several injection points in one page, but each of them is
distinct. Now, the filter is only looking for one pair of brackets, so
the attacker can't fool the extension into leaving part of the
injection unsanitized. However, they could, for example, inject an
extra ending bracket after the opening parenthesis (or an extra
starting brace). The extension will be tricked into sanitizing
non-malicious content, the black pluses (+). This behavior can be
detected by noting that we know the order in which the
\code{endPoints} should appear, and so if the filter sees a closing
\code{endPoint} before the next expected starting \code{endPoint}, or
similarly, a starting \code{endPoint} before the next expected closing
\code{endPoint}, this attack can be identified. In the diagram, the
order of the solid elements characterizes the possible malformations
in the end points. As with the previous scenario, we have to sanitize
the outermost end points, potentially deleting non-malicious
content. The signature developer specifies the sanitization behavior.

Note that these complex cases do not mean that our approach is not always applicable. The process of writing the signature might become more complicated, but the extension provides a choice for blocking the page entirely if the signature writer believes a given case is too complex for our signature language.

\subsection{Dynamic injections} \label{dynamic_injections}

The top-level documents of web pages fetch additional dynamic content
via \js{fetch} or AJAX APIs. Content fetched in
this way is also vulnerable to \xss, and must be filtered. An example
vulnerability is CVE-2018-7747 (WordPress Caldera Forms, which allows malicious
content retrieved from the plugin's database to be injected in response to a click.

\sys allows XHR requests to be filtered with \js{xhr}-type
signatures. To reduce the number of signatures that need to be
considered when a browser issues a request, we require that signatures
for XHR be nested inside a signature for a top-level document. If a
page's main content matches an existing top-level signature description,
\sys will then enable all nested XHR listeners.

Signatures for dynamic requests are specified in the \js{listenerData}
key, which includes a listener type and method. The idea is extensible
to scripts and other objects loaded separately from the main document
(e.g., images, stylesheets, etc.).

\begin{lstlisting}[breaklines=true,caption={
      An example dynamic request signature. This patches CVE-2018-7747.
    },label={lst:dynamic_signature}]
...
listenerData: [{
  listenerType: 'xhr', listenerMethod: 'POST',
  sanitizer: 'escape', type: 'string',
  listenerUrl: 'wp-admin/admin-ajax.php',
  typeDet: 'single-unique',
  endPoints: ['<p><strong>', '[AltBody]']
}]
\end{lstlisting}

%% file: implementation.tex
\section{Implementation} \label{implementation}

We implemented our system as an extension in Firefox 69.0. Our signatures are stored in a local JavaScript file in the extension package. We decided on an extension implementation for several reasons.
(1) \emph{Privileged execution environment}. The extension's logic lies in a separate environment from the web application code. This guarantees that malicious code in the application cannot affect the extension. 
(2) \emph{Web application context}. Our solution requires knowledge of the application's context. 
The extension naturally retains this context.
(3) \emph{Interposition abilities}. As it lies within the browser, the extension can run both at the network level, e.g., rewrite an incoming response; and at the web application level, e.g., interpose on the application's JavaScript execution. 

\begin{algorithm}[tb]
	\SetAlgoLined
	//global DBSignatures\\
	\SetKwProg{verifyResponse}{procedure \emph{verifyResponse}}{}{end}
	\verifyResponse{(responseString, url)}{
	 loadedProbes = runProbes(responseString, url)\\
	 signaturesToCheck $\leftarrow$ []\\
	 \For{probe in loadedProbes}{
	 	signaturesToCheck.append(DBSignatures[probe])\
 	 }
  	 filteredSignatures $\leftarrow$ []\\
  	 \For{signature in signaturesToCheck} {
  	 	\uIf{responseString and url match signature}{
  	 		filteredSignatures.push(signature)\\
  	 	}
  	 }
     versionInfo $\leftarrow$ loadVersions(url, loadedProbes)\\
     endPoints $\leftarrow$ [] \\
     \For{signature in filteredSignatures} {
     	\uIf{(signature,signature.version) $\in$ versionInfo}{
     		endPoints.push(signature.endPointPairs)\\
     	}
    }
	indices $\leftarrow$ [] \\ 
	\For{endPointPair in endPoints} {
		indices.push(findIndices(responseString, endPointPair))\\
	}
	\uIf{discrepancies exist in indices}{
       	Block page load and return\\
	}
	\For{endPointPair in endPoints} {
		sanitize(responseString,indices)\\
	}
		\caption{Network filter algorithm}
		\label{filter_algorithm}
	}
\end{algorithm}

\subsection{Filtering process} \label{filtering_process}
\autoref{filter_algorithm} describes our network filtering process: once a request's response comes in through the network, we process it and sanitize it if necessary.

\textbf{Loading signatures}
Our detector loads signatures and finds injection points in the document. However, not all signatures need to be loaded for a specific website, since not all sites run the same frameworks. When loading signatures, we proceed in a manner similar to a decision tree. The detector first probes the page (line 3) to identify the underlying framework (the \textbf{software} in our signature language). We currently provide a number of static probes. However, as more applications are required to be included, we believe it would be better to cover this task in the signature definitions. The widely popular network mapping tool Nmap~\cite{nMap} uses probes in a similar manner, kept in a modifiable file. As mentioned in \autoref{viability}, we currently only have signatures for CMS applications. Our probes use specific identifiers related to the application, as well as the particular site that is affected by the exploit. WordPress pages, for example, have several elements in the page that identify it as a WordPress page. While this might seem easier for CMS style pages, and we acknowledge that application fingerprinting is a hard problem in general, we believe other web apps will also have similar identifying information, like headers, element ID's, script/CSS sources, etc.

After running these probes, the detector loads corresponding frameworks' signatures and filters out checks whether the information of each loaded signature matches the page (lines 5-12).

\textbf{Version identification}
We then apply version identification (lines 13-16). Our objective for versioning is that our signatures don't trigger false positives on websites running patched software. We found this to be one of the harder aspects of signature loading. In many \acp{CMS}, for example, file names are not updated with the latest versions, or do not include them at all, and thus, this information is often hard to come by from the client-side perspective. This information is often more available to admins of a site. While this might not be the bulk of users, it is the bulk of disclosed CVEs, as described in \autoref{viability}.

Furthermore, we believe that even if we load a signature when the application has already been fixed at the server-side, it will often preserve the page's functionality, as many of the CVEs are a result of unsanitized input. Motivated by this observed behavior, our mechanism follows a series of increasingly accurate but less applicable version identifiers: first, we apply framework specific version probes. If these are not successful, the signature language provides functionality for version identification in the HTML through regex. If information is unavailable through the HTML, the version in the signature is left blank and the patch is applied regardless of version, as we can not be sure the page is running patched software. Our tool takes advantage of having perfect knowledge of an exploit's conditions, which reduces the rate of false positives compared to a software-agnostic approach.

\textbf{Injection point search and sanitization}
Once we have the correct signatures, we find the indices for the endpoints using our top-down, bottom-up scan, and need to check for potential malformations in the injection points, as described in \autoref{multiple_injections} (lines 19-24). The page load is blocked and a message is returned to the user, or if the signature developer specifies so, sanitization proceeds on the new endpoints. Finally, if all \textbf{endPoint} pairs are in the expected order, we sanitize each injection point (lines 25-27).

\subsection{Sanitization methods}
We provide different types of sanitization: "DOMPurify", "escape", and "regex". Regex Pattern matching can be particularly effective when the expected value has a simple representation (e.g., a field for only numbers). For each of these approaches, the signature can specify a corresponding config value. DOMPurify provides a rich API for additional configuration. When escaping, defining specific characters to escape via regex can be useful. For pattern matching via regex, config specifies the value the injected should match.

%% file: signature_development.tex
\section{Writing Signatures} \label{signature_writing}

We expect a signature developer to have a solid understanding of the principles behind \ac{XSS}, as well as web applications, HTML, CSS and JavaScript, so they can identify precise injection points. In this section, we aim to show that minor effort is required from a knowledgeable analyst when writing a signature.

\subsection{Case Study: CVE-2018-10309} \label{case_study}
Going back to our example in \autoref{motivating_example}, we describe the process for writing a signature using one of our studied CVEs.

\textbf{Identifying the exploit.} An entry in Exploit Database
\cite{studyCVE} describes a persistent \ac{XSS} vulnerability in the
WordPress plugin Responsive Cookie Consent for versions
1.7/1.6/1.5. This entry (as most do) comes with a \ac{PoC} for the
exploit, which describes the Cookie Bar Border Bottom Size parameter
as vulnerable. We run a local WordPress installation with this plugin. In general, the system does not rely on the existence of a \ac{PoC}, we personally relied on this as we did not discover the CVEs and did not have the full context of the exploit.

\textbf{Establishing the separation between dynamic and static content.} We insert the string "$>$script$>$alert('XSS')$<$/script$>$ in the Cookie Bar Border Bottom Size (rcc\_settings[border-size] in the HTML) input field. This results in an alert box popping up in the page.

In general, the analyst is able to find the vulnerable HTML from the server-side code without having to reproduce the exploit. Since we did not write the CVE, we had to do this extra step.

In the example, the \textbf{input} element is the injection starting point, and the \textbf{label} tag is the end point, since it is the element immediately after the \textbf{input}. Identification of correct endpoints is extremely important, and in particular, when a page has multiple injection points, the signature developer must ensure the chosen elements do not overlap with other innocuous ones. In some cases, the developer might think it best to completely stop the page from loading. While one of our main goals is to maintain the page’s usability, there are cases where large portions of the document would be affected by the sanitization. We believe compromising usability for security is preferable in this case. Furthermore, the developer has to identify if the exploit comes from an external source (such as an Ajax request), as this changes the signature.

\textbf{Collecting other required page information and writing the signature.} The next step is to gather the remaining information to determine whether the signature applies to the page loaded. The full signature for this example was previously shown in \autoref{lst:xsnare_signature}. The \textbf{URL} is acquired by noting that this exploit occurs on the plugin's settings page. The \textbf{software} running is WordPress in this case. The settings page's HTML includes a link to a stylesheet with href "http://localhost:8080/wp-content/plugins/responsive-cookie-consent...", in particular, "wp-content/plugins/plugin-name" is the standard way of identifying that a WordPress page is running a certain plugin, in this case, "responsive-cookie-consent", set as \textbf{softwareDetails}. We apply the signature for all versions less than or equal to 1.7. Since the exploit only occurs in this specific spot in the HTML, the \textbf{typeDet} is listed as "single-unique". 
Since the vulnerable parameter is a border-size, the \textbf{sanitizer} applied is "regex", further restricting the pattern to only numbers in \textbf{config}. We list the \textbf{endPoints} as taken from the HTML.

\textbf{Testing the signature}. Finally, we load up our extension and reload the web page. We expect to not have an alert box pop up, and we manually look at the HTML to verify correct sanitization. In practice, there might be small discrepancies between server-side and client-side representations of the HTML string, leading to bugs in the signature if the developer used the parsed HTML as a reference. If the exploit is not properly sanitized, the developer is able to use the debugging tools provided by the browser to check the incoming network response information seen by the extension's background page and make sure it matches the signature values.

%% file: methodology.tex
\section{Approach evaluation} \label{viability}

To verify the applicability of our detector and signature language, we tested the system by looking at several recent CVEs related to \ac{XSS}. We have three objectives: to verify that our signature language provides the necessary functionality to express an exploit and its patch, to test our detector against existing exploits, and to show that composing signatures takes a reasonable amount of time.


\subsection{Methodology} \label{methodology}

We study recent CVEs related to WordPress
plugins. We focus on WordPress for two reasons:
\begin{enumerate}
	\item WordPress powers 34.7\% of all websites according to a recent survey  \cite{w3stats} \cite{DBLP:journals/corr/abs-1801-01203}. The same study states that 30.3\% of the Alexa top 1000 sites use WordPress. Thus, we can be confident that our study results will hold true for the average user.
	\item WordPress plugins are popular among developers (there are currently more than 55,000 plugins \cite{wpplugins}). Due to its user popularity, WordPress is also heavily analyzed by security experts. A search for WordPress CVEs on the Mitre CVE database \cite{cvemitre} gives 2310 results. Plugins, specifically, are an important part of this issue, 52\% of the vulnerabilities reported by WPScan are caused by WordPress plugins \cite{wpscan}.
\end{enumerate}

We used a CVE database, CVE Details~\cite{cvedetails} to
find the 100 most recent WordPress \ac{XSS} CVEs, as of
October 2018. 
%
For each CVE, we set up a Docker container with a clean installation
of WordPress 5.2 and installed the vulnerable plugin's version. 
For CVEs that depended on a particular WordPress version, we
installed the appropriate version. Of the CVEs we looked at, only
one occurred in WordPress core. We believe it would be harder
to precisely sanitize injection points in WordPress core, as many of
the plugins have particular settings pages where the exploits occur,
and the HTML is more identifiable. WordPress core, on the other hand,
can be heavily altered by the use of themes and the user's own
changes. However, as evidenced by our investigation, the vast majority
of exploits occur in plugins.
 
Next, we reproduced the exploit in the CVE and we analyzed the
vulnerable page and wrote a signature to patch the exploit.

\subsection{Results}

\begin{table}[h!]
	\begin{center}
		\begin{tabular}{c c} 
			\hline
			\textbf{Plugin} & \textbf{Installations}\\ [1ex] 
			\hline
			WooCommerce  & 5+ million  \\  
			Duplicator & 1+ million \\  
			Loginizer & 900,000+ \\  
			WP Statistics & 500,000+ \\  
			Caldera Forms & 200,000+ \\   
			\hline
		\end{tabular}
		\caption{Most popular studied WordPress plugins}
		\label{table:1}
	\end{center}
\end{table}

Of the initial 100 CVEs, we were able to analyze 81 across 44 affected
pages. We dropped 24 CVEs due to reproducibility issues: some of the
descriptions did not include a PoC, making it difficult for us to
reproduce; or, the plugin code was no longer available. In some cases,
it had been removed from the WordPress repository due to "security
issues", which emphasizes the importance of being able to defend
against these attacks. 

The resulting plugins we studied averaged 489,927
installations: \autoref{table:1} shows the number of installations for
the 5 most popular plugins we studied. For the vulnerabilities, 27
(35.5\%) could be exploited by an unauthenticated user; 56 (73.7\%)
targeted a high-privilege user as the victim, 7 (9.2\%) had a
low-privilege user as the victim, the rest affected users of all
types.

Many of the studied CVEs included attacks for which there are known
and widely deployed defenses. For example, many were cases of
Reflected \ac{XSS}, where the URL revealed the existence of an attack,
e.g.,: \url{http://<target>&page-uri=<script>alert("\ac{XSS}")</script>}
While Chrome's built-in \ac{XSS} auditor blocked this request, Firefox
did not, and so we still wrote signatures for such attacks\footnote{In
practice, we found several cases where even XSS auditor did not block
a reflected XSS.}.

We wrote 59 WordPress signatures in total, which got rid of the PoC
exploit when sanitized with one of our three methods. Note that while
a PoC is often the most simple form of an attack, our sanitization
methods, and in particular DOMPurify, can get rid of complex
injections as well. We were able to include several CVEs in some PoCs
because they occurred in the same page and affected the same
plugin. Overall, these signatures represent 71 (93.4\%) signed
CVEs. The 5 we were not able to sign were due to lack of identifiers
in the HTML, which would result in potentially large chunks of the
document being replaced\footnote{In these cases, the signature developer
can weigh the trade-offs and decide whether the added cost is worth
it.}

After manual testing, the majority of the 71 signatures maintained the
same layout and core functionality of the webpage. However, 12
signatures caused some elements to be rearranged, modifying the page's
visual aspect. One caused a small part of the page to become unusable,
due to the sanitization method (a table showing user information
was now rendered as blank). Most of the responsibility of maintaining
functionality is left to the signature developer. We found that being precise
is key to retaining functionality.

While our goal is to retain as much information of the page as
possible after sanitization, we believe that even if a part of the
page becomes unusable, this does not impact the user's experience as
much, since many of the exploits occur in small sections of the
HTML. A usability study is out of scope for this paper and we leave it
to future work.


\subsection{Generalizability beyond WordPress}
\label{generalizability}

To test the generalizability of our approach to other frameworks, we
analyzed 5 additional CVEs, 2 related to Joomla!, 2 for LimeSurvey,
and 1 for Bolt CMS.  We chose Joomla! because it is another
popular \ac{CMS}. Unfortunately, we only found 2 CVEs
that we were able to reproduce, as the software for its extensions is
often not available. For fairness, we looked for the most recent CVEs
we could reproduce listed in the Exploit Database~\cite{exploitdb}, since
these have recorded \acp{PoC}. We carried out the same
procedure as with the WordPress CVEs, and were able to patch all of
the 5 exploits. This brought our CVE coverage rate up to 94.2\%.

\subsection{Signature writing times} \label{signature_times}

\begin{figure}[h]
	\begin{center}
	\includegraphics[width=\linewidth]{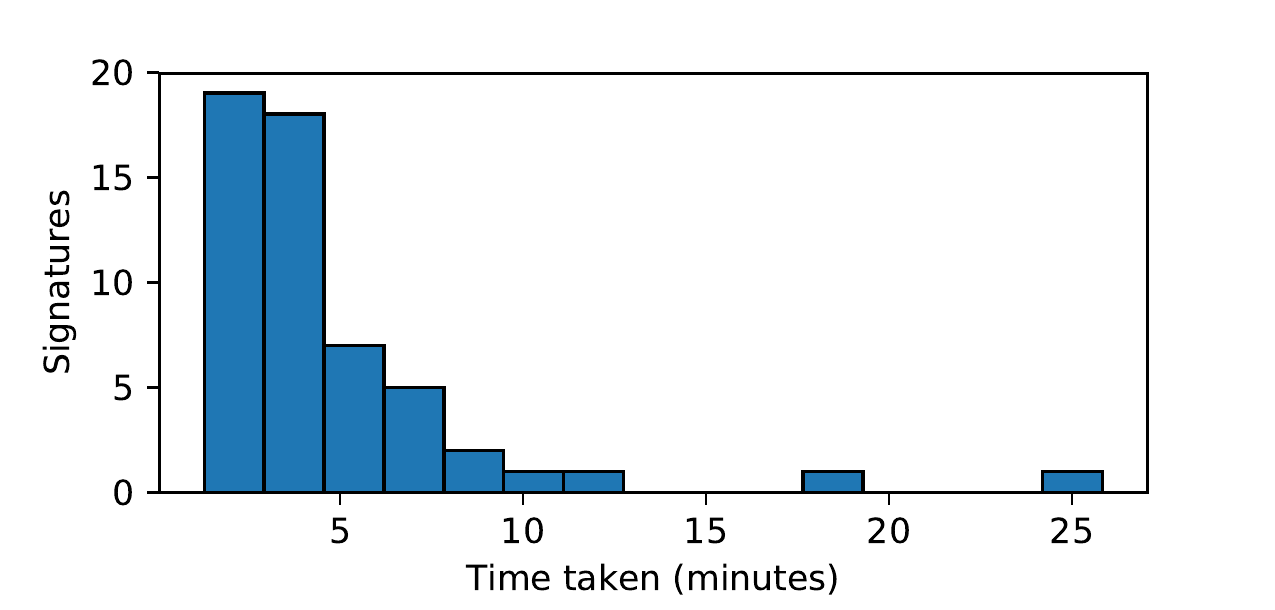}
	\caption{Histogram of time taken to write signatures.}
	\label{fig:signature_times}
	\end{center}
\end{figure}

\autoref{fig:signature_times} plots a histogram of the times it took
one of the authors to compose each of the signatures. Each time
measurement includes the time it took to check the HTML injection
points, write the signature and to debug it. We do not include the
time taken to discover and carry out an exploit, as this is part of
the CVE writing process. The median time is 3.89 minutes, and the
standard deviation is 4.18 minutes. 72\% of signatures were written in
under 5 minutes. We believe this to be a reasonable amount of time
considering the security granted by our extension.

The signature which took the longest time to write (25 minutes)
corresponds to an exploit with 12 HTML injection points. Additionally,
testing this signature proved difficult, as some of
the injections were a result of a script inserting elements in the DOM
after the page had loaded. This caused the initial HTML to look
innocuous, but with exploits still occurring after sanitization. As
this script was part of the initial request, we eventually got to the
root of the problem. We believe a more experienced exploit analyst
might be able to detect this kind of behaviour more
easily. 

The signature which took the second longest time to write (18 minutes)
corresponds to an exploit with 7 injection points; each of
these belongs to a part of the HTML with generic element
identifiers. Our language provides a means to overcome this, by allowing
the developer to specify the element's "position": for example, if there are
three <h3 class="title"> elements, in the HTML, and only one of them
is an injection starting point, the writer can specify that the third
one needs to be sanitized. The same can be done for the ending points. As
there were 7 of such points, debugging took longer than for
other signatures.

Another source of longer timings are complicated 
listener type signatures, like the ones for exploits caused by an
XHR. We had 4 such signatures and composed these with a median time of 9.86 minutes.
As only a small number of these were present, we expect the time
taken to compose a typical signature to be lower than the overall median in this experiment.

%% file: performance.tex
\section{Load time performance on top websites} \label{performance}

\sys's performance goal is to provide its security guarantees without
impacting the user's browsing experience. We now briefly report \sys's
impact on top website load times, representing the expected behaviour
of a user's average web browsing experience. For more performance
evaluation results please see ~\autoref{appendix:perf-eval}.

%
For these tests we used the top 500 websites as
reported by Moz.com~\cite{top500}. For each website, we loaded it 25
times (with a 25 second timeout) and recorded the following values:
requestStart, responseEnd, domComplete, and decodedBodySize. From the
initial set of 500, we only report values for 441: the other 59 had
consistent issues with timeouts, insecure certificates, and network
errors. In our setup, we used a headless version of Firefox 69.0, and
Selenium WebDriver for NodeJS, with GeckoDriver. All experiments were
run on one machine with an Intel Xeon CPU E5-2407 2.40GHz processor,
32 GB DRAM, and our university's 1GiB connection.

We ran four test suites:
\textbf{No extension cold cache}: Firefox is loaded without the extension installed and the web driver is re-instantiated for every page load.
\textbf{Extension cold cache}: As before, but Firefox is loaded with the extension installed.
\textbf{No extension warm cache}: Firefox is loaded without the extension installed and the same web driver is used for the page's 25 loads.
\textbf{Extension warm cache}: As before, but Firefox is launched with the extension installed.

For each set of tests, we reduced the recorded values to two comparisons: network filter (responseEnd - requestStart), and page ready (domComplete - responseStart). The first analyzes the time spent by the network filter, while the second determines the time spent until the whole document has loaded. We calculate the medians for each website for each of these measures as well as the decodedByteSize.

\begin{figure}[h]
		\begin{center}
	\includegraphics[scale=0.5]{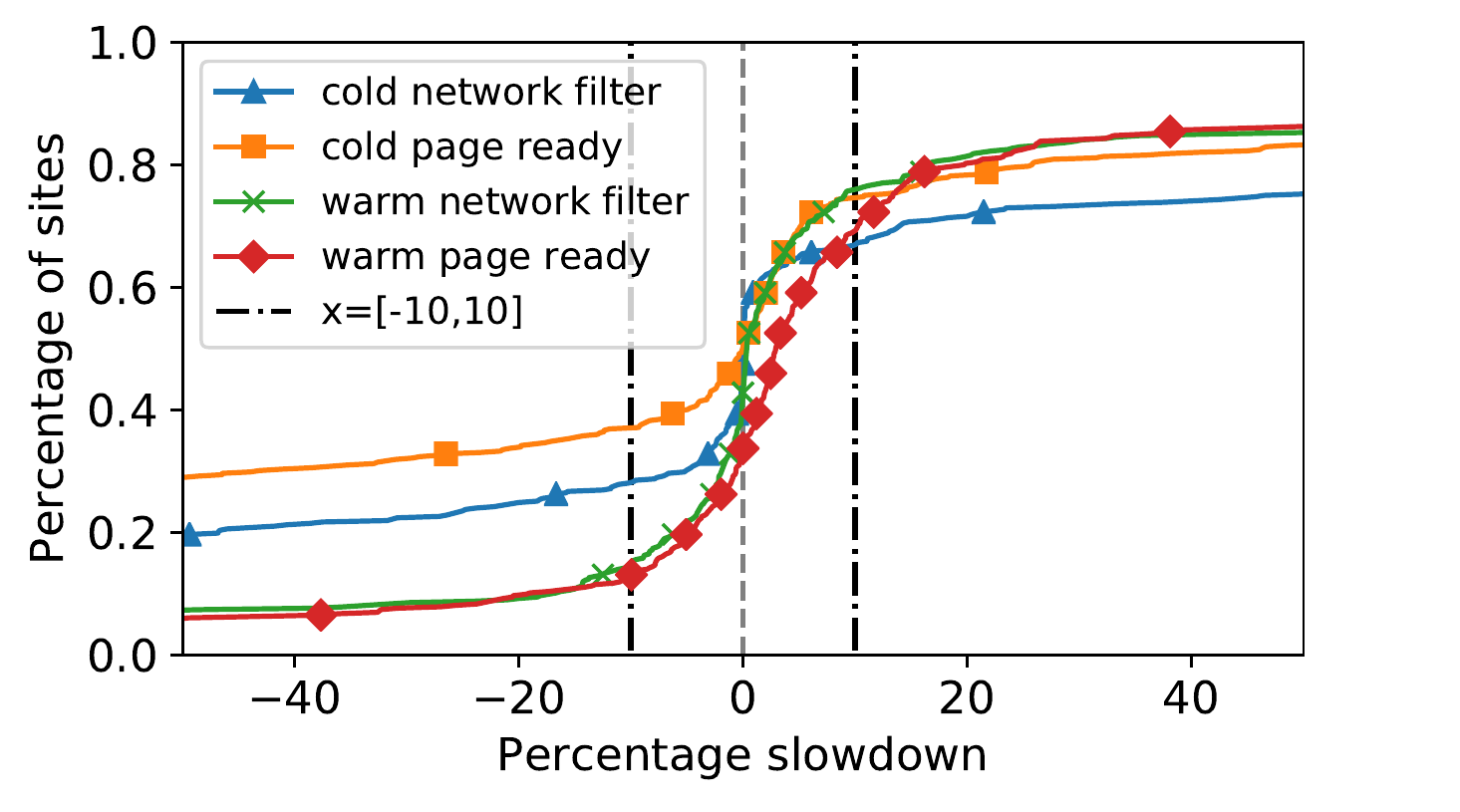}
	\caption{Cumulative distribution of relative percentage slowdown with extension installed for top sites.}
	\label{fig:overall_slowdown}
\end{center}
	
\end{figure}

We compare the load times with/without the extension by calculating the relative slowdown with the extension installed according to the following formula:

\begin{equation*}
100*\frac{\tilde{x}_{with}-\tilde{x}_{without}}{\tilde{x}_{without}}
\end{equation*}
\\
where $\tilde{x}$ is the median with/without the extension running.

\autoref{fig:overall_slowdown} plots the results. The graph
shows a slowdown of less than 10\% for 72.6\% of sites, and less than
50\% for 82\% of sites when the extension is running. Note that these
values are recorded as percentages, and while some are as high as
50\%, the absolute values are in 77\% of cases less than a
second. This overhead should not alter the user's experience
significantly.

The slowdown increases by at most 5\% when we take caching into
account. This is likely because the network filter causes the browser
to use less caching, especially for the DOM component, as it might
have to process it from scratch every time. While it may seem
counter-intuitive that some pages have shorter loading times with
the extension, there are several variables at play that can affect
these measurements (local network, server-side load, internal
scheduling, etc). We manually checked the websites for which values
were higher than |40\%| and verified that our extension did not change
the page's contents, a possible cause of faster load times. We also
checked the timings for the page as reported by the browser and noted
a high variance even within small time windows. The time spent by our
verification function was less than 10ms for 87.6\% of sites
(\autoref{fig:verification_time_string_length}). This corroborates our
findings that the slowdown is mostly negligible.

%% file: limitations.tex
\section{Limitations and Future Work}

\textbf{Generalizability and scope of study.} As discussed in \autoref{methodology}, while many websites share similar structures to the ones we covered, our study only considered 4 other sites apart from those running on WordPress, and we only considered sites using a CMS. Not all websites might be identified as easily. Furthermore, we only studied 81 CVEs. 
In the future we intend to study a more diverse set of CVEs and go beyond CMS-based sites.

\textbf{False positives and false negatives.} Due to the nature of our approach, it is extremely hard to completely get rid of FPs: If the applied sanitization targets JavaScript code, for example, a FP will likely be triggered. Furthermore, since we rely on handwritten signatures to defend against attacks, vulnerable sites for which no signature has been written will be subject to FNs. In the future, we intend to study the rate of FPs and FNs in our approach and compare it to previous work.


\textbf{Usability.} A main aspect of our work is its increased potential for usability and adoption from both a user's perspective that installs the extension to defend themselves against \ac{XSS}, and a signature developer who has to write the database descriptions according to a known CVE. Future work could focus on usability studies related to both of these aspects.

\textbf{Protection against CSRF.} We believe that we can adapt our work to defend against \ac{CSRF} exploits, as well. Using a similar signature language as the one for \ac{XSS}, a signature developer could specify pages with potential vulnerabilities to only allow network requests that cannot exploit such vulnerabilities.


\textbf{Dealing with an increasing number of signatures.} As the number of framework probes increases, and more types of sites are covered, the performance impact will increase. Using more efficient approaches to searching and filtering, and using better data structures in the signature database could to lower this overhead.


\textbf{Design considerations}. Currently, each browser user has to install our extension. However, the same functionality could be offloaded to a single processing unit similar to a proxy, which can handle the filtering for all machines in a network. This deployment model might be more appropriate in certain environments, such as in enterprises.



%% file: related_work.tex
\section{Related Work}
We classify existing work into several categories: client-side, server-side, browser built-in, and hybrid: a combination of these.

\noindent \textbf{Server-side techniques.}
In addition to existing parameter sanitization techniques,
taint-tracking has been proposed as a means to consolidate
sanitization of vulnerable parameters~\cite{Xu:2006:TPE:1267336.1267345,DBLP:conf/sec/Nguyen-TuongGGSE05,Pietraszek:2005:DAI:2146257.2146267,Bisht:2008:XPD:1428322.1428325}. These
techniques are complementary to ours, and provide an additional line
of defence against \ac{XSS}. However, many of them rely on the
client-side rendering to maintain the server-side properties, which
will not always be the case.

\noindent \textbf{Client-side techniques.}
DOMPurify~\cite{10.1007/978-3-319-66399-9_7} presents a robust
\ac{XSS} client-side filter. The authors
argue that the DOM is the ideal place for sanitization to occur. While
we agree with this view, their work relies on application developers
to adopt their filter and modify their code to use it. Thus, we have partly automated this step by including it as our default sanitization function.

Jim et al.~\cite{Jim:2007:DSI:1242572.1242654} present a method to
defend against injection attacks through Browser-Enforced Embedded
Policies. This approach is similar to ours, as the policies specify
prohibited script execution points. However, this again relies on application developers knowing where their code might be vulnerable. Furthermore, browser modifications are required to benefit from it. Similarly, Hallaraker and Vigna~\cite{Hallaraker:2005:DMJ:1078029.1078861} use a
policy language to detect malicious code on the client-side. Like \sys, they make use of signatures to protect against known types of
exploits. However, unlike our approach, their signatures are not
application-specific, and there is no model for signature
maintenance. Furthermore, there is no evaluation on the efficacy of
their signatures.

Snyder et al.~\cite{Snyder:2017:MWD:3133956.3133966} report a study in which
they disable several JavaScript APIs and test the number of websites
that are do not work without the full functionality of the APIs. This approach increases security due to vulnerabilities present in several
JavaScript APIs, however, we believe disabling API functionality
should only be used as a last resort.

Similarly to server-side defences, taint-tracking has been applied at the client-side: DexterJS provides a robust, browser-independent platform for auto-patching DOM-based XSS \cite{10.1145/2786805.2786821,10.1145/2786805.2803191}. While this approach effectively defends against a large number of attacks automatically, it only covers a subset of possible XSS attack. This applies to any client-side defence that is unaware of an application's server-side code.

\noindent \textbf{Browser built-in defences.}  Browsers are equipped
with several built-in defences. We previously described XSS
Auditor in \autoref{introduction}, another important one is the 
 \ac{CSP}~\cite{CSP}. It has been widely adopted and
in many cases provides developers with a reliable way to protect
against \ac{XSS} and \ac{CSRF} attacks. However, \ac{CSP} requires the developer to identify which scripts
might be malicious.

\noindent \textbf{Client and server hybrids.}
XSS-Dec~\cite{Sundareswaran:2012:XHS:2352970.2352994} uses a proxy which keeps track of an encrypted version of the server's source files, and applies this information to derive exploits in a page visited by the user. This approach is similar to ours, since we assume previous
knowledge of the clean HTML document. Furthermore, they use anomaly-based and signature-based detection to prevent attacks. However, there is no mention of signature maintenance. In a way, our system offloads all this functionality to the client-side, without the need for any server-side information.

%% file: conclusion.tex
\section{Conclusion}

Users cannot depend on administrators to patch vulnerable server-side
software or for developers to adopt best practices to mitigate XSS
vulnerabilities. Instead, users should protect themselves with a
client-side solution. In this paper we described the design,
implementation, and evaluation of \sys, one such client-side approach.
\sys prevents \ac{XSS} exploits by using a database of exploit
signatures and by using a novel mechanism to detect XSS exploits in a
browser extension.
We evaluated \sys through a study of 81 CVEs in which we
showed that it defends against 94.2\% of the exploits.



%% file: appendix.tex
\appendix

\section{Performance evaluation} \label{appendix:perf-eval}

In this section we report additional performance measurements for \sys.

\textbf{Methodology.} We recorded timestamps while our code is
executing using the Performance Web API\footnote{Note that while this
  API normally reports values as doubles, due to recent security
  threats, such as Spectre~\cite{DBLP:journals/corr/abs-1801-01203},
  several browser developers have implemented countermeasures by
  reducing the precision of the DOMHighResTimeStamp
  resolution~\cite{reducetimeprecision,resolutionconsiderations}. In
  particular, Firefox reports these values as integer
  milliseconds. For our tests, we re-enabled higher precision values.}

\begin{figure}[h]
	\begin{center}
 \includegraphics[scale=0.65]{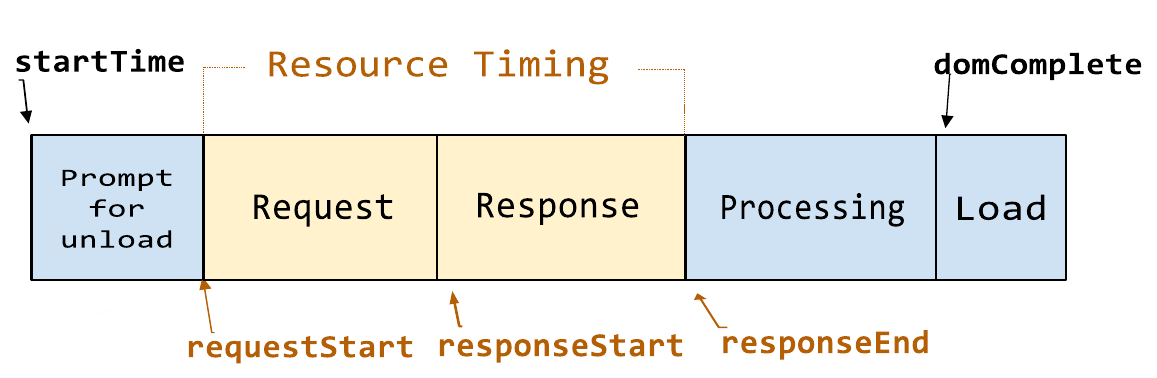}
 \end{center}
 \caption{The Navigation Timing API's timestamps\protect\footnotemark}
 \label{fig:navigationtiming}
 \end{figure}

While our extension's functionality only applies at the network level,
there is potential slowdown at the DOM processing level due to the
optimization techniques the browser applies throughout several levels
of the web page load pipeline. \autoref{fig:navigationtiming} shows
the different timestamps provided by the Navigation Timing
API~\cite{navigationtiming}, as well as a high-level description of
the browser processing model. Since our filter listens on the
onBeforeRequest event, none of the previous steps before Request are
affected. In this section, we refer to the difference in time between
responseEnd and requestStart as the "network filter time".

\footnotetext{This image was taken from the w3 spec: \url{https://www.w3.org/TR/navigation-timing-2/}}

\subsection{Top websites load times; continued} \label{top_sites}

\begin{figure}[h]
	\includegraphics[scale=0.5]{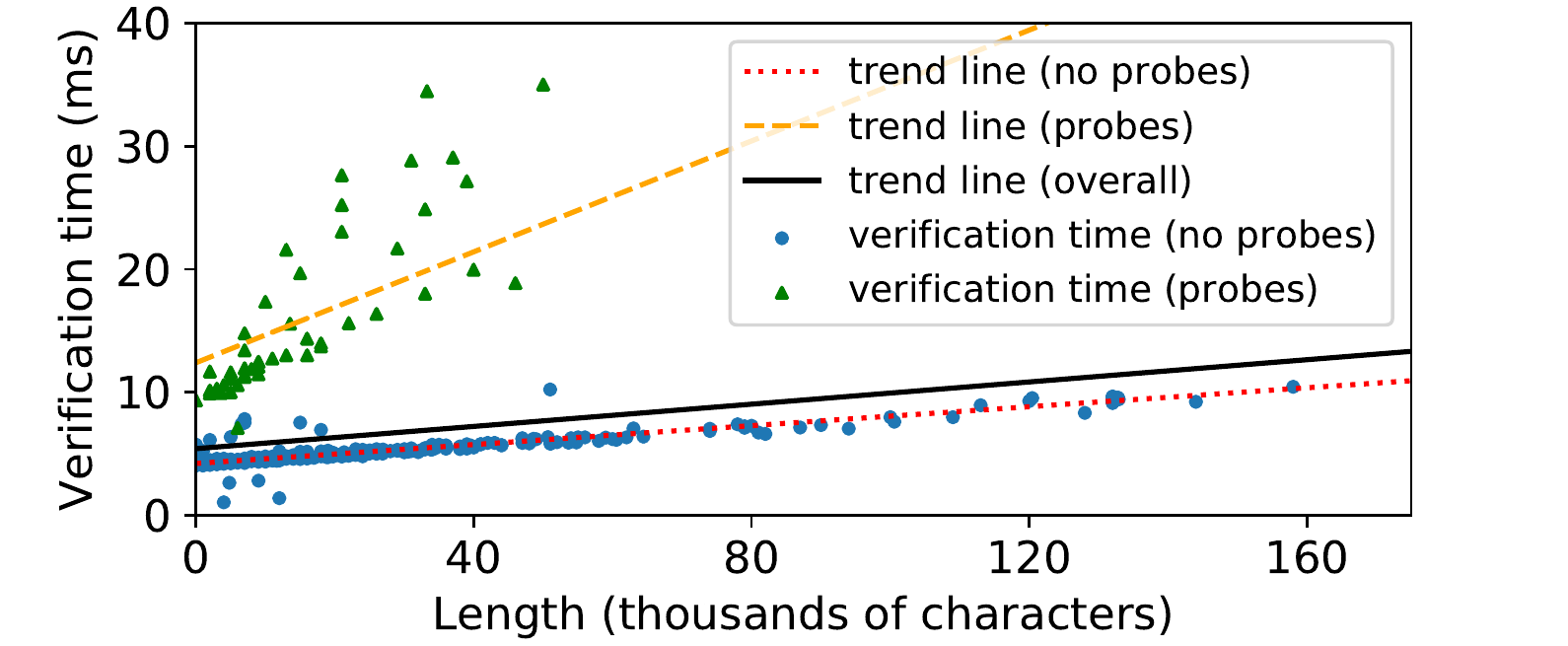}
	\caption{Scatter plot of network filter time as a function of character length for top sites.}
	\label{fig:verification_time_string_length}
\end{figure}

 \autoref{fig:verification_time_string_length} shows the time spent by the call to our string verification function in the network filter as a function of the length of the string to be verified. The blue dots are the pages for which our framework probes tested negative, and the green triangles are the pages for which the probes tested positive: 55 in total. We applied least squares regression to calculate the shown trend lines. The Spearman's rank \footnote{The Spearman's correlation coefficient measures the strength and direction of association between two ranked variables: https://statistics.laerd.com/statistical-guides/spearmans-rank-order-correlation-statistical-guide.php} correlation values for no probe, probe, and overall are 0.91, 0.91, and 0.72 respectively, demonstrating positive correlation. Since both our probes and signatures use regex matching, we expect both trend lines to be linear, as seen in the graph. Recall that once a probe for a certain software passes, we perform a linear scan over the signatures for that specific software and check whether it applies to the given HTML string or not. Thus, we expect the slope of the line to be higher when a probe passes. Around 37.4\% of all web sites use frameworks covered by our probes~\cite{w3stats}, thus, we expect the impact of our network filter to be closer to the non-probe values, as corroborated by our overall trend line.

\textbf{False positives on the Web.} Additionally, for each website, we recorded the number of loaded signatures. We report a 0\% FP rate for loaded signatures. Thus, we can infer with confidence that the rate of false positives for loaded signatures during an average user's web browsing is similarly low. This rate could possibly go up as the number of signatures and covered frameworks increases. It is likely that these websites are free of vulnerabilities covered by our signatures, as many of these websites are not running WordPress to begin with, and being the most popular, they would likely be updated quickly if a vulnerability is found; thus, the rate of false negatives is likely extremely low as well.

\subsection{WordPress websites load times} \label{wordpress_sites}

We ran similar experiments as in Section 6.1, but with the WordPress sites described in Section 4.1. Thus, all of these have either one or multiple injection points in their HTML, and the network filter will spend an additional amount of time sanitizing these as defined by the signatures. Note that the data set is smaller here, and some of the trends might be harder to infer.

\begin{figure}[h]
	\includegraphics[scale=0.5]{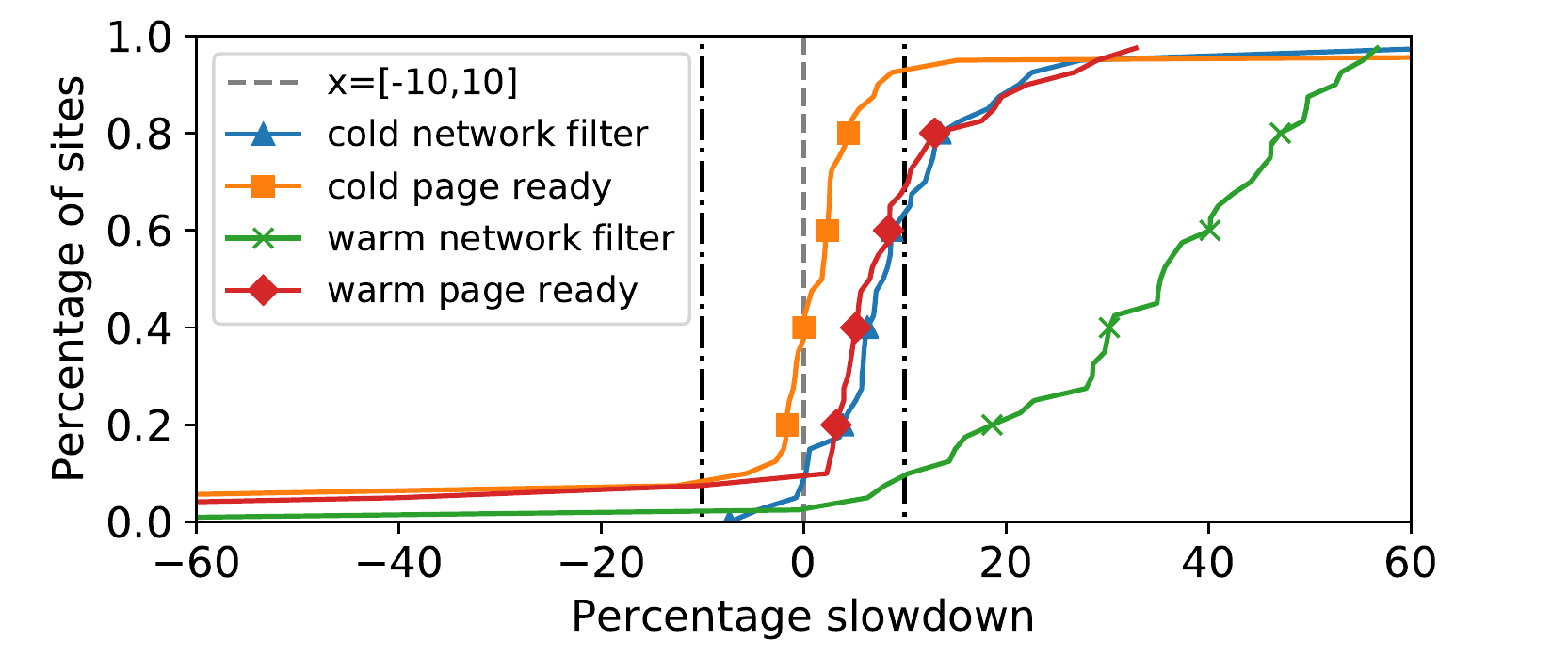}
	\caption{Cumulative distribution of relative percentage slowdown with extension installed for WordPress sites.}
	\label{fig:WordPress_slowdown}
\end{figure}

\autoref{fig:WordPress_slowdown} shows the results for slowdown with the extension running for these sites. Recall that the only difference between a page which passes the WordPress probe and one that matches a signature is that the latter has to replace a portion of the original string by its sanitized version. In this case we see a slowdown of less than 10\% for 60\% of sites, and less than 40\% for 96.25\% of them. The warm network filter curve suffers from a particularly high slowdown. We believe this to be the case because the locally hosted pages decrease the network component time, causing any overhead to be seen as relatively high. However, as 48\% of the original values were below 60ms, conclude a small impact on user experience as well.

\begin{figure}[h]
	\begin{center}
	\includegraphics[scale=0.55]{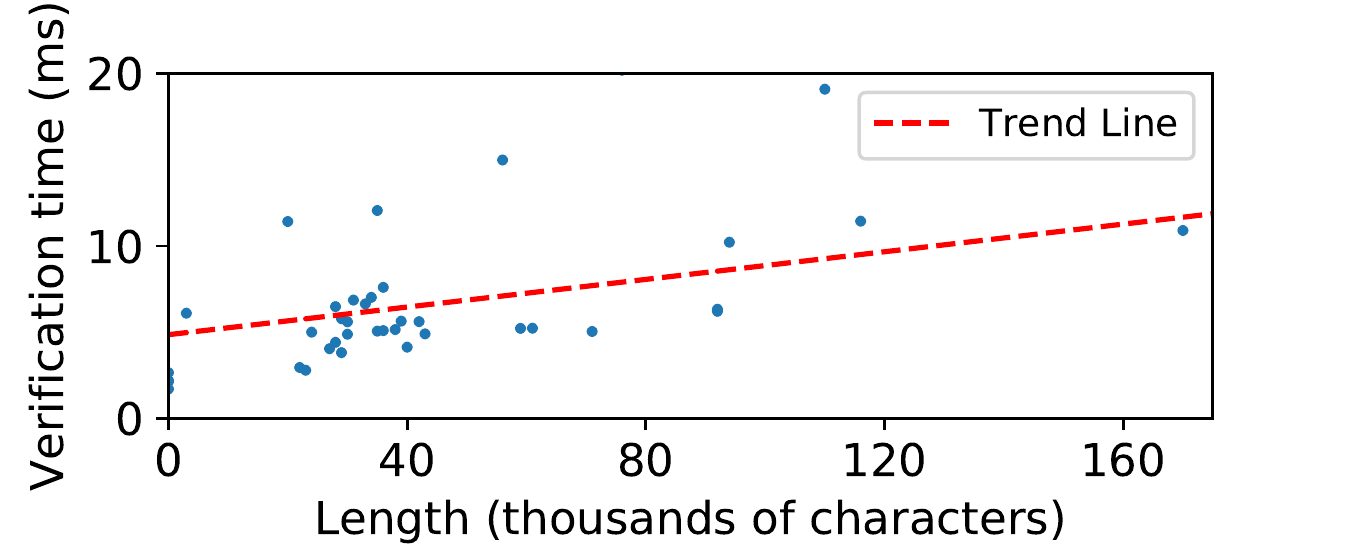}
	
	\caption{Scatter plot of network filter time as a function of character length for WordPress sites.}
	\label{fig:verification_time_string_length_wordpress}
\end{center}
\end{figure}

Finally, we report the string verification time as a function of its length, for the WordPress sites, shown in \autoref{fig:verification_time_string_length_wordpress}. The Spearman's rank correlation for this set of data is 0.630.

\section{Signature Language Specification} \label{appendix:language_specification}

We provide a description of our signature language, in particular in the context of WordPress:
\begin{itemize}
	\item
	url: If the exploit occurs in a specific URL or subdomain, this is defined as a string, e.g.
	\url{/wp-admin/options-general.php?page=relevanssi\%2Frelevanssi.php}, otherwise null.
	\item
	software: The software framework the page is running if any, e.g. WordPress. A hand-crafted page
	might not have any identifiable software.
	\item
	softwareDetails: If running any software, this provides further information about when to load a signature. For WordPress, these are plugin names as depicted in the HTML of a page running such plugin.
	\item
	version: The version number of the software/plugin/page. This is used for versioning of the software run by the page, as described in \autoref{filtering_process}.
	\item 
	type: A string describing the signature type. A value of "string" describes a basic signature. A value of 'listener' describes a signature which requires an additional listener in the background page for network requests.
	\item 
	sanitizer: A string with one of the following values: "DOMPurify", "escape", and "regex". This item is optional, the default is DOMPurify.
	\item
	config: The config parameters to go along with the chosen sanitizer, if necessary. For "DOMPurify", the accepted values are as defined by the DOMPurify API (i.e, DOMPurify.sanitize(dirty, config). For "escape", an additional escaping pattern can be provided. For "regex", this should be the pattern to match with the injection point content.
	\item
	typeDet: A string with the following pattern: 'occurrence-uniqueness', 'ocurrence' has values single/multiple, which describes the existence of one or multiple independent injection points; the 'uniqueness' has values unique/several, specifying whether an injection point occurs once or several times throughout the document, as described in Section 2.4.
	\item
	endPoints: An array of startpoint and endpoint tuples, specified as strings for regex matching.
	\item 
	endPointsPositions: An array of integer tuples. These are optional but useful when the one of the endPoints HTML are used throughout the whole page and appear a fixed number of times. For example: if an injection ending point happens on an element <h3 class='my-header'>, this element might have 10 appearances throughout the page. However, only the 4th is an injection ending point. The signature would specify the second element of the tuple to be 7, as it would be the 7th such item in a regex match array (using 1-based indexing), counting from the bottom up. For ending points, we have to count from the bottom up because the attacker can inject arbitrarily many of these elements before it, and vice versa for starting points.
\end{itemize}

Additionally, if the value of type is `listener', the signature will have an additional field called listenerData. Similarly to a regular signature, this consists of the following pieces of information:
\begin{itemize}
	\item 
	listenerType: The type of network listener as defined by the WebRequest API (e.g. `script', `XHR', etc.)
	\item
	listenerMethod: The request's HTTP method, for example "GET" or "POST".
	\item
	url: the URL of the request target.
\end{itemize}
If a listener is present, the signature's fields can be used to specify the listener's request injection points.